\documentclass{article}                    % onecolumn
\usepackage{graphicx}
\usepackage{amsmath,amssymb}  \usepackage{eufrak}
\usepackage{mathptmx}      % use Times fonts if available on your TeX system
\usepackage{xcolor}
\begin{document}

\title{A dual first-postulate basis for special relativity}

%%\titlerunning{A dual first-postulate basis for special relativity}        % if too long for running head

\author{Brian Coleman}

%\authorrunning{Brian Coleman} % if too long for running head
%
%\institute{Brian Coleman---BC Systems (Erlangen) \at
%             Velchronos, Ballinakill Bay\\ Moyard, County Galway, Ireland\\
%             % Tel.: +353-86-8133327\\
%              \email{brian.coleman.velchron@gmail.com}  }

\maketitle

\noindent \emph{BC Systems (Erlangen)}\\   
\emph{Velchronos, Ballinakill Bay\\ Moyard, County Galway, H91HFH6, Ireland}\footnote{brian.coleman.velchron@gmail.com}\\

\begin{abstract}\footnote{http://iopscience.iop.org/article/10.1088/0143-0807/24/3/311 \\ Note: Already published corrigenda have now been incorporated and typos rectified. Footnote references 5 and 10, and a  MAPLE solution weblink for  \S 5's single clock measurement formulae, are added in this posting. Notably, \S 6's derivations have been more directly established in a 2017 book [36].} 
\noindent An overlooked straightforward application of velocity reciprocity to a triplet of inertial frames in collinear motion identifies the ratio of their cyclic relative velocities' sum to the negative product as a cosmic invariant---whose inverse square root corresponds to a universal limit speed. A logical indeterminacy of the ratio equation establishes the repeatedly observed unchanged speed of stellar light as \emph{one instance} of this universal limit speed. This formally renders the second postulate redundant. The ratio equation furthermore enables the limit speed to be quantified---in principle---\emph{independently} of a limit speed signal. Assuming negligible gravitational fields, two deep-space vehicles in \emph{non-collinear} motion could measure with only a single clock the limit speed against the speed of light---\emph{without requiring these speeds to be identical}. Moreover, the cosmic invariant (from dynamics, equal to the mass-to-energy ratio) emerges \emph{explicitly} as a function of signal response time ratios between three \emph{collinear} vehicles, multiplied by the inverse square of the velocity of whatever arbitrary signal might be used. \end{abstract}

Published 2003 in~European Journal of Physics Vol.24 pp.301-313

%\vspace*{2cm}
\section{First-postulate foundations}
\noindent Special relativity theory is traditionally established on the basis of the relativity postulate---the equivalence of inertial frames---together with Albert Einstein's postulation of the constancy of the velocity of light. It is not widely appreciated that this `second postulate' as such inappropriately anchors special relativity in the domain of electromagnetism and, moreover, is redundant. The primarily mathematical approaches of previous papers on this topic however have been considered as `not available for general education' [23].\footnote{Mermin's paper also discusses measuring the limit speed using the triplet velocity equation, and mentions that a limit speed signal is in principle not required.} 

\begin{figure}[t]  
{\centering \includegraphics[width=67mm,height=49mm]{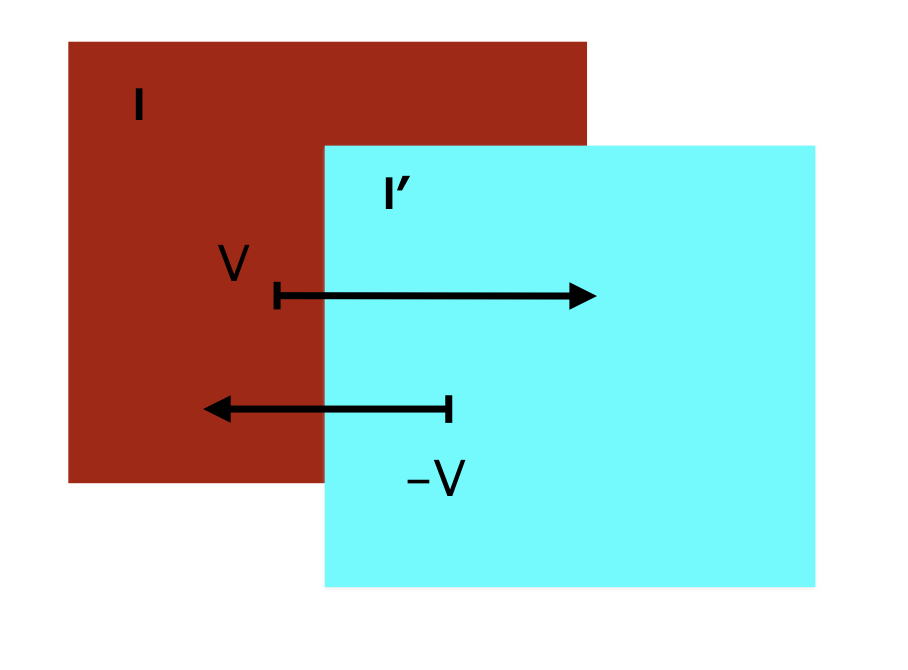} \par} 
\caption{Two inertial frames in relative motion.}\label{fig1}  
 \end{figure}

Using Henri Poincar\'{e}'s relativity principle\footnote{Einstein's [8] and Poincar\'{e}'s [6, 7] parallel evolutions [14] of special relativity had been preceded by several attempts to detect the `ether wind': by Michelson [1] in Potsdam near Berlin (1881) and (with Morley) in Cleveland (1887), by FitzGerald/Trouton [4] in Dublin (1901)---a virtually unknown epochal event---and by Trouton/Noble [5] in London (1903).} [6, 7], `all inertial frames experience the laws of physics in an equivalent manner', we assume \emph{spatial isotropy, spatial homogeneity, time homogeneity, velocity reciprocity} and \emph{uniqueness of compound velocity}. Consequently all inertial frames have identical standards of length and of time progression and we may restrict our initial considerations to a pair of single-dimension inertial frames $I$ and $I'$ , with $I'$ moving at fixed velocity $v$ as perceived from $I$. (Figure 1. Diagram arrows indicate the velocity of the frame containing the arrow head, as perceived by the frame containing the arrow tail.)

Any event can be pinpointed in either frame by the distance and time $I$-frame coordinates $x$, $t$ or $I'$-frame coordinates $x'$, $t'$ respectively. On coincidence of the frames' origins $I_o$ and $I'_o$ , both frame times may be for convenience set to zero.

As is well known, directly from spatial and temporal uniformity we can then relate $x$ in terms of $x'$ and $t'$ and, conversely, $x'$ in terms of $x$ and $t$, by similar linear equations, where $F_{v}$ , $G_{v}$, $F_{-v}$ and $G_{-v}$ are four initially unknown parameters which are however dependent solely on the interframe velocities $v$ and $-v$ respectively:

\begin{equation} x'=x/F_{v}+t/G_{v}\qquad \textup{and}\qquad x=x'/F_{-v}+t'/G_{-v} .\end{equation}
Likewise familiar scenarios involving mutual perception of origin trajectories and unit-length rods then easily resolve three of these unknown parameters: $1/G_v =-v/F_v$, $1/G_{-v} =v/F_{-v}$ and $F_{-v} = F_v$.
This reduces (1) to
\begin{equation}  x'=\frac{x-tv}{F_v}\qquad \textup{and}\qquad x=\frac{x'+t'v}{F_v} .\end{equation}
Hence
\begin{equation} t'=\frac{t-x(1-F_v^2)/v}{F_v}\end{equation}
Noting the symmetry between ((2)i) and (3), we define a new term---chronocity---(not to be
confused with the medical term `chronicity'), as a factor corresponding to \emph{distance-rated time `displacement'}:
\begin{equation} \textup{Chronocity } \kappa_v\equiv \frac{1-F_v^2}{v}.  \end{equation}
From (2)-(4) we can then relate $x'$ and $t'$ respectively in terms of $x$ and $t$, \emph{symmetrically}:
\begin{equation} x'=\frac{x-tv}{F_v}\qquad \textup{and}\qquad t'=\frac{t'-x\kappa_v}{F_v} . \end{equation}
The single remaining unknown parameter $F_v$ we label \emph{the FitzGerald contraction factor} [2, 4, 33]. $F_v$ is normally established by invoking the second postulate. This conventionally entrenched approach however---as we now show in a rather simple manner---is quite unnecessary.
\begin{figure}[t]  
{\centering \includegraphics[width=90mm,height=55mm]{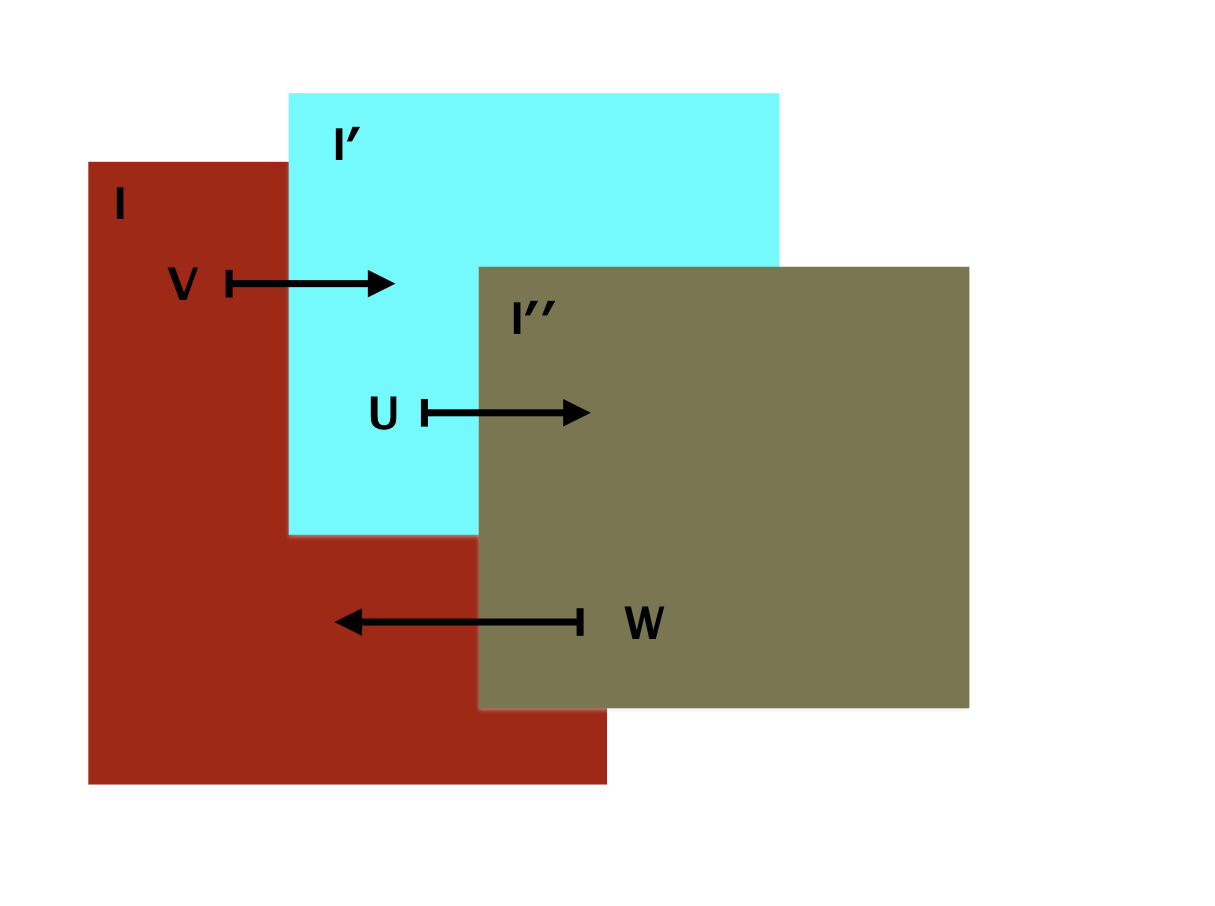} \par} 
\caption{A triplet of inertial frames.}\label{fig2}  
 \end{figure}

\section{The emergence of a cosmic constant}
\subsection{A hitherto unnoticed simple physical path}
\footnote{A shorter \emph{physical} path to equations [10-12] below, appears in [36].}We introduce (figure 2) a third frame $I''$ with coordinates $x''$, $t''$, and its origin $I''_o$---also''coincidentally synchronized with $I_o$ and $I'_o$---perceived by $I$ to be travelling in the positive $x'$-direction at velocity $u$, collinear with $v$. The velocity of $I$ as perceived by $I''$ we denote by $w$ whose orientation is chosen for \emph{cyclic symmetry}.
\newpage
The equations for $x'', t''$ in terms of $x',t'$ depend only on $u$ and are of course identical to (5), with chronocity $\kappa_u \equiv (1 - F_u^2)/u$:

\begin{equation} x''=\frac{x'-t'u}{F_u}\qquad \textup{and}\qquad t''=\frac{t''-x'\kappa_u}{F_u} . \end{equation}
Next we derive $x''$ and $t''$ in terms of $x$ , $t$ , $v$ and $u$ (and dependent contraction and chronocity factors $F_v$, $F_u$, $\kappa_v$ and $\kappa_u$), eliminating $x'$ and $t'$ by substituting ((5)i) and ((5)ii) in ((6)i) and ((6)ii):

\begin{equation} x''=\frac{x(1+u\kappa_v)-t(v+u)}{F_v F_u}\qquad \textup{and}\qquad t''=\frac{t(1+v\kappa_u)-x(v+u)}{F_v F_u}. \end{equation}
Equations similar to (7) were used by Terletskii [16] and Rindler\footnote{Rindler's 1990 [25] and 2001 [34] books did not refer to his 1969 derivation [17] which commented on [9] and [12] with the remark `However, like numerous others [papers] that followed these have gone largely unnoticed'.  Rindler's 1969 derivation was kindly brought to my attention by John Stachel, author of [21] and [31].} [17] who however then invoked abstract group theory and required the context of a two-postulate derivation, instead of using the following direct \emph{physical} argument (surprisingly also missed by others such as [9-13,15-25,29,30,32,35]): the velocity of origin $I_o$ ($x = 0$), as perceived by $I''$, is $w = x''/t''$. \\ 
To obtain $x''/t''$ we use ((7)i) and ((7)ii):

\begin{equation} x''=\frac{0\cdot(1+u\kappa_v)-t(v+u)}{F_v F_u} \qquad \textup{and}\qquad t''=\frac{t(1+v\kappa_u)-0\cdot(v+u)}{F_v F_u} .  \nonumber \end{equation}
Therefore
\begin{equation} w=\frac{x''}{t''} =\frac{-(v+u)}{1+v\kappa_u} .\end{equation}
The velocity of $I''$ ($x'' = 0$), as perceived by $I$, is $-w = x/t$. Equation ((7)i) gives
\begin{equation} 0=\frac{x(1+u\kappa_v)-t(v+u)}{F_v F_u}  \nonumber\end{equation}
Hence
\begin{equation} -w=\frac{x}{t} = \frac{v+u}{1+u\kappa_v}  \end{equation}
From (8) and (9),
\begin{equation}  \frac{v+u}{1+u\kappa_v} = \frac{v+u}{1+v\kappa_u} . \nonumber\end{equation}
Therefore

\begin{equation}  \frac{\kappa_v}{v} = \frac{\kappa_u}{u} \triangleq \Omega  .   \end{equation}
Since $v$ and $u$ are arbitrary, the above \emph{chronocity/velocity} ratios---independent of one another but still equal---must be \emph{invariant}. We are thus presented with \emph{a new universal constant}---$\Omega$--- with dimension inverse velocity squared. 
\newpage
\noindent This corresponds to [9] what Vladimir Ignatowski---a Russian scientist born in Georgia in 1875---first described in a lecture in Moscow in December 1909:
\begin{quote}
On the basis of the relativity principle alone, it is possible to prove that a universal cosmic constant must exist, in contrast to [the method of] Einstein, who in parallel with the relativity principle assumes a priori the velocity of light as a universal constant. In proving the existence of the above constant, we will refer not at all to the speed of light, and derive the existence of this constant in a general sense, and not on the basis of any special physical phenomenon. (Present author's translation.)
\end{quote}
\begin{figure}[t]  
{\centering \includegraphics[width=46mm,height=66mm]{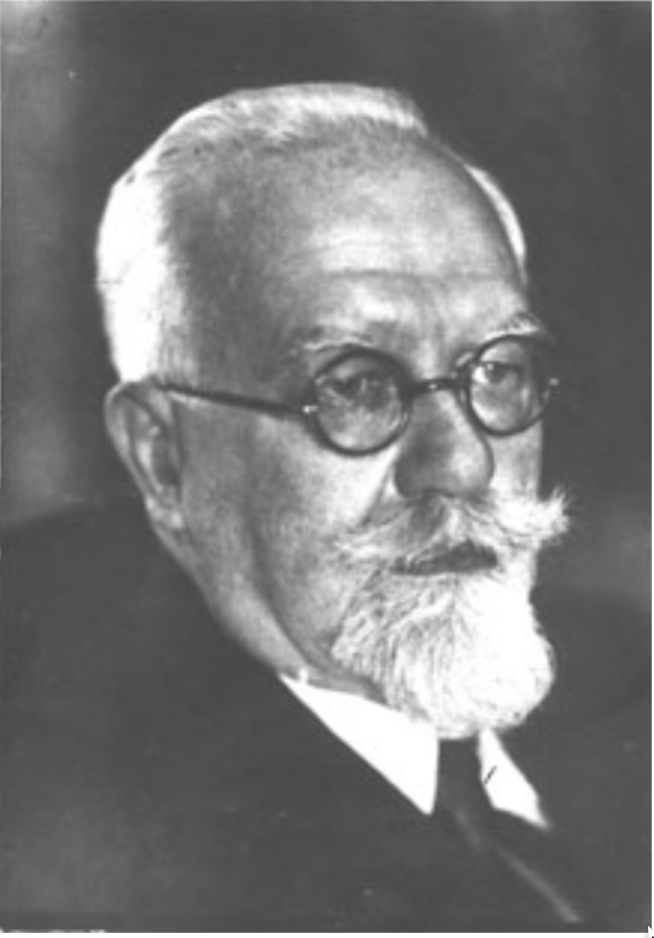} \par} 
\caption{Vladimir Ignatowski (1875Ð1942).\newline Photograph courtesy of
the General Physics Institute, Moscow (see Acknowledgments).}\label{fig:Ignat}  
 \end{figure}
Ignatowski was executed in 1942 during the siege of Leningrad, having been falsely accused of spying for the Germans. He was posthumously rehabilitated in 1955 [27].\footnote{This information was kindly provided in translated form by Sergei Yakovlenko of the Moscow General Physics Institute (see Acknowledgments).}

\noindent Equations (9) and (10) immediately give us \emph{Einstein's equation for velocity addition}:

\begin{equation} -w=\frac{v+u}{1+vu\Omega}.\end{equation}
This can be expressed in a more general and useful cyclic form as \emph{the triplet velocity ratio equation}:
\begin{equation} \Omega=\frac{v+u+w}{-vuw}.\end{equation}
\newpage
\begin{figure}[t]  
{\centering \includegraphics[width=90mm,height=66mm]{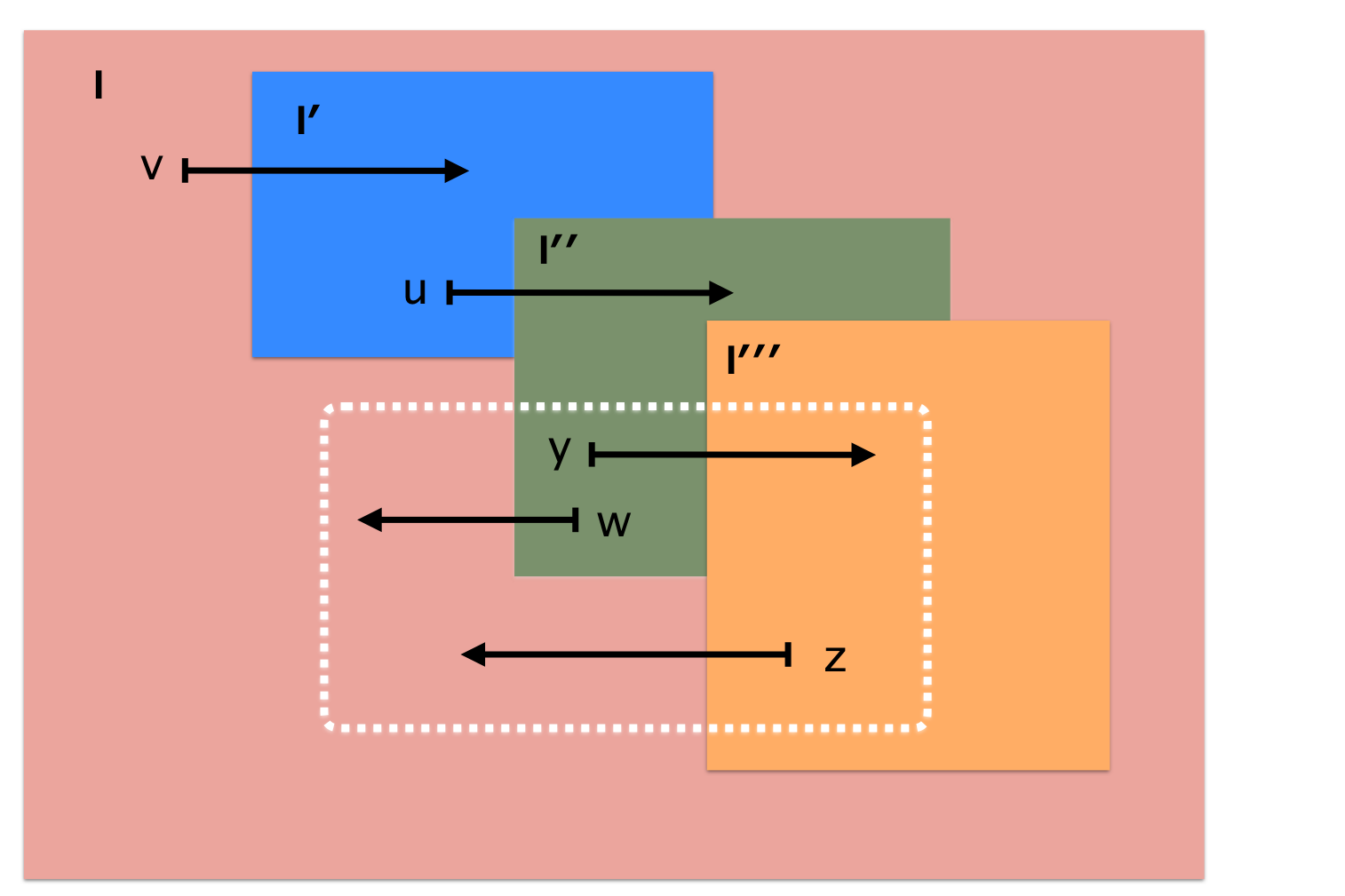} \par} 
\caption{Four frames.}\label{fig4}  
 \end{figure}
 \vspace*{-10mm}
\subsection{$\Omega$'s connection with a cosmic limit speed}
Adding (figure 4) a fourth frame $I'''$ with velocities $y$ and $-z$ relative to frames $I''$ and $I$ respectively and applying (12) to the frame triplet $I$, $I''$ and $I'''$, we obtain
\begin{equation} \Omega=\frac{-w+y+z}{wyz}\qquad \textup{i.e.} \qquad  w=\frac{y+z}{1+yz\Omega}  \nonumber  \end{equation}
Substituting for $w$ in (12) gives us
\begin{equation} \Omega=\frac{v+u+y+w}{-(vuy+uyz+yzv+zvu)}. \end{equation}
If we make $u$ and $y$ both equal to $v$ in (13) and consider $-z$, the velocity of $I'''$ perceived by $I$, then
\begin{equation} \Omega=\frac{3v+z}{-v^3-3v^2z } \qquad \textup{i.e.} \qquad   -z=\frac{3v(1+\Omega v^2/3)}{1+3\Omega v^2}.     \end{equation}
If $\Omega$ is negative, then at $v = 1/\sqrt{-3\Omega}$ , $-z$ is infinite. Higher-order equi-recessive cascades would reduce this singularity threshold to any desired lower proportion of whatever $1/\sqrt{-\Omega}$  might be (as mentioned by Rindler [17]); i.e.
\begin{equation}      \emph{Velocity addition singularity excludes any finite negative value for  }  \Omega. \end{equation}

Returning to three frames (figure 2) and putting $u = v$ in (11), gives
\begin{equation} -w=\frac{2v}{1+\Omega v^2}. \nonumber \end{equation}
%\vspace*{-4mm}
This peaks at $v =1/\sqrt{\Omega}$, which means that any particular non-zero $-w$-value could result from two different $v$-values, one below and one above $1/\sqrt{\Omega}$. 

\newpage
\begin{figure}[t]  
{\centering \includegraphics[width=90mm,height=50mm]{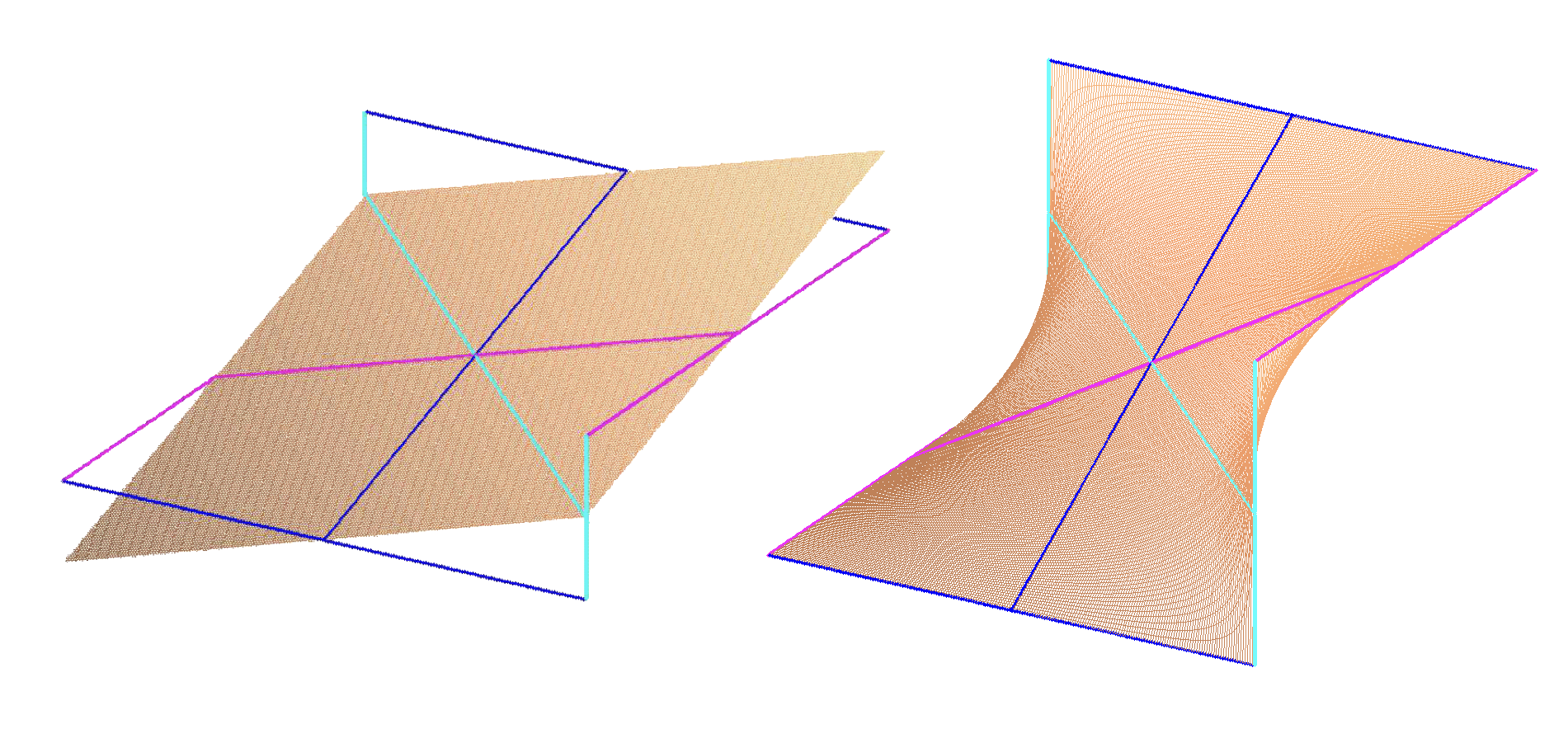} \par} 
\caption{ Zero- and non-zero-omega triplet velocity surfaces.}\label{fig5}  
 \end{figure}
\vspace*{-7mm}
\noindent Hence uniqueness also excludes possible interframe velocities above $1/\sqrt{\Omega}$. (Field [32] likewise used uniqueness to reach this conclusion.)
Consequently,
\begin{align}  & \emph{Velocity addition uniqueness implies that } 1/\sqrt{\Omega} \nonumber \\ & \emph{ corresponds to a cosmic upper limit speed}. \end{align} 

This cosmic limit speed $1/\sqrt{\Omega}$ is infinite or finite in accordance with whether $\Omega$  turns out to be zero or positive non-zero. These two cases are illustrated (figure 5) by the two triplet velocity surfaces in $vuw$-space defined by the triplet velocity equation (12). In the zero-$\Omega$ case, the surface is a flat plane of infinite extent which corresponds to classical physics velocity addition. In the non-zero-$\Omega$  case we restrict our attention to the region of velocities within the range $\pm1/\Omega$   allowed by velocity addition uniqueness. 

Noteworthy on the non-zero surface are the three spatially tri-symmetrical `H' figures (one
is shown with dashed lines), each of which traces the points where a particular frame perceives (or is perceived by) the other two frames to have the same equal but cyclically opposite velocity, e.g.\ where the velocity of $I$ is the same from $I'$ as from $I''$ (i.e.\ $w = -v$). From equation (12) we then have $u - v^2u\Omega  = 0$ which factors as $u(1+v\sqrt{\Omega})(1-v\sqrt{\Omega}) = 0$. 
Each `H' crossbar (e.g.\ $u = 0$) represents the situation where one frame views the speeds of the other two frames to be equal but below $1/\sqrt{\Omega}$, with the latter frames being mutually stationary (as on the zero-$\Omega$ surface also). 

The `H' sidebars (e.g. $w = -v =\pm 1/\Omega )$---which traverse the non-zero-$\Omega$  surface only---describe where both of these latter frames are perceived to have the positive or negative limit speed respectively,but they may have an arbitrary speed ($u$) relative to each other, in stark contrast to the classical physics case. From this we may draw two further significant conclusions:
\vspace*{-1mm}
\begin{align} 
 & \emph{If two inertial frames observe a third to have the same collinear speed which is} \nonumber \\  & \emph{below the cosmic limit speed } 1/\sqrt{\Omega}, \emph{ the two frames are mutually stationary}. \end{align}   
\vspace*{-4mm}
\begin{align} 
 & \emph{ If two inertial frames in actual relative motion perceive a third to have the same} \nonumber \\  & \emph{collinear speed, this speed is the cosmic limit speed} 1/\sqrt{\Omega}. \end{align}   

\subsection{The Lorentz transformation equations and the concept of chronocity}
Equation (10) directly resolves our chronocity factor:

\begin{equation} \kappa_v=v\Omega \end{equation}
From (4) and (10) we obtain $\Omega$  as a function of $F_v$ :

\begin{equation} \Omega=\frac{1-F_v^{2}}{v^2}\end{equation}
and conversely \emph{the FitzGerald contraction factor}:

\begin{equation} F_v=\sqrt{1-v^2\Omega}\end{equation}

Equations (5), (19) and (21) then give {the Lorentz transformation equations}\footnote{Irish-born Joseph Larmor, professor at Galway and Cambridge (and friend of FitzGerald), was the first to refer to `local time' and (in 1898) to exactly formulate what were subsequently designated by Poincar\'{e} as the \emph{Lorentz transformation equations}.} [3], with $\Omega$ as yet unquantified:

\begin{equation} x'=\frac{x-tv}{\sqrt{1-v^2\Omega}} \qquad \textup{and} \qquad  t'=\frac{t-xv\Omega}{\sqrt{1-v^2\Omega}}  .\end{equation}

Considering two distinct but \emph{complementary} scenarios where either $x' = 0$ or $t' = 0$ in the Lorentz transformations (22), we have either $x/t = v$ or $t/x = v\Omega$ respectively. The \emph{origin displacement per unit time} (velocity $v$) is thus the {distance/time separation ratio} in frame $I$ between two events \emph{space-coincident} in frame $I'$ (e.g.\ $x ' = 0$). Similarly our chronocity factor $\kappa_v = v\Omega$  constitutes the \emph{time/distance separation ratio}, i.e.\ \emph{disparity of simultaneity per unit distance} in frame $I$, between two events which are {time-coincident} in frame $I'$ (e.g.\ $t' = 0$). Chronocity is accordingly the \emph{space–time counterpart} to velocity $v$ in (22); i.e.,

\begin{equation} x'=\frac{x-tv}{\sqrt{1-v\kappa_v}} \qquad \textup{and} \qquad  t'=\frac{t-x\kappa_v}{\sqrt{1-v\kappa_v}}  .\end{equation}

\begin{figure}[t]  
{\centering \includegraphics[width=90mm,height=50mm]{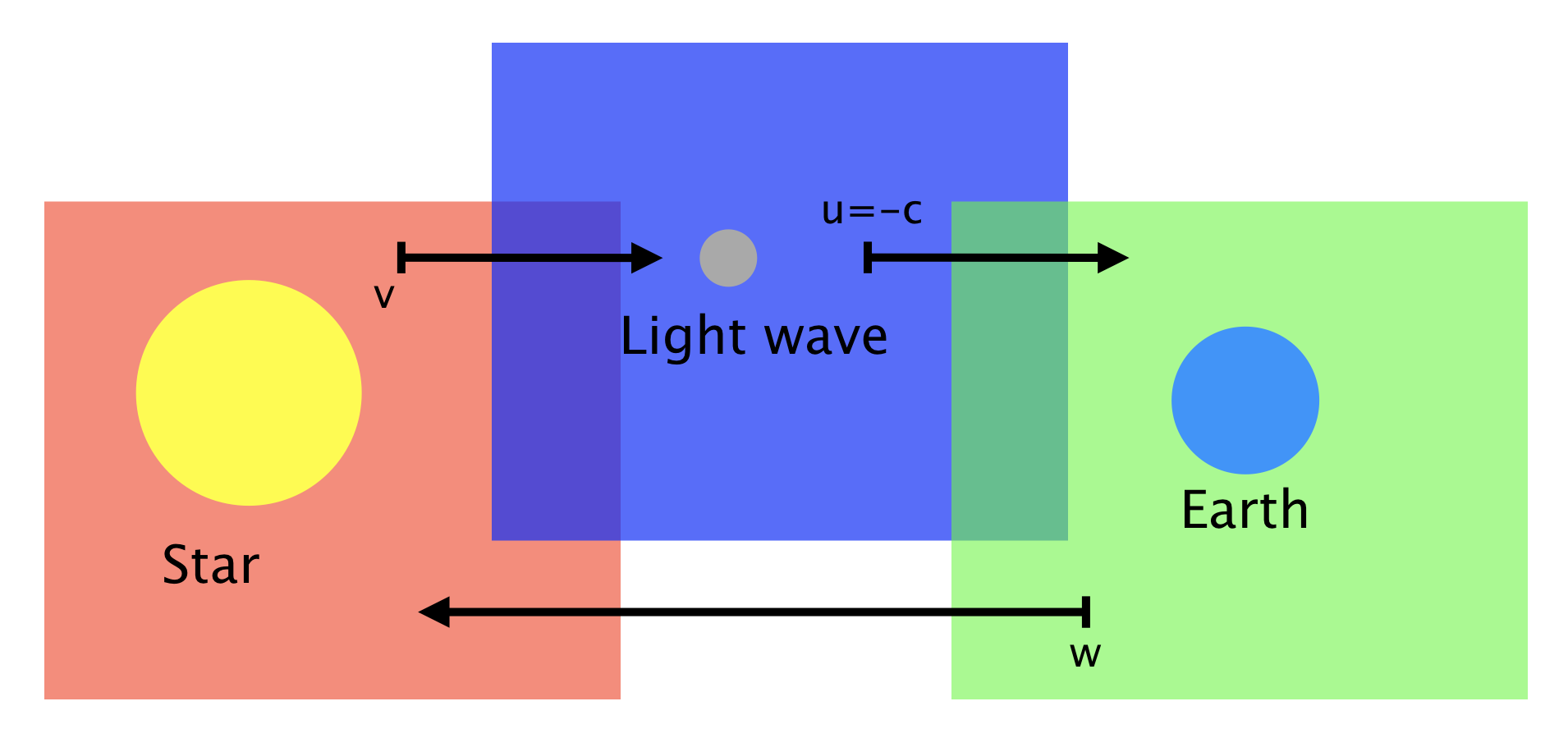} \par} 
\caption{A star/light wavefront/Earth frames triplet.}\label{fig6}  
 \end{figure}

We can thereby express this space–time symmetry, which reflects the pivotal logic (10) originally identifying the invariant, as a kernel statement of special relativity:
\begin{align} 
 & \emph{Interframe chronocity equals interframe velocity} \nonumber \\  & \emph{divided by the cosmic limit speed squared.} \end{align}

\section{Quantification of $\Omega$ by virtue of indeterminacy}
Just as the \emph{independence} of one quotient (10) led to the birth of our cosmic constant, another property of another quotient is a convenient key to its coincidental quantification---that of \emph{indeterminacy}.

Velocities ($u$) relative to Earth of light, i.e. electromagnetic waves emitted by stars at unknown relative velocities ($v$), were noted already in 1729 by James Bradley, from aberration observations, to appear to have the same value $-c$, regardless of whatever velocities ($w$) the individual stars might have relative to Earth (shown schematically in figure 6).

Putting $u = -c$ in (11) gives

\begin{equation} -w=\frac{v-c}{1-cv\Omega} .\end{equation}
Without taking for granted \emph{a priori} that $v$ has any consistent value, we can assume however that in \emph{at least two cases} among numerous observations $w$ differs significantly. The multivalued ratio

\begin{equation} \frac{v-c}{1-cv\Omega}    \nonumber \end{equation}
therefore---being independent even of its only conceivably varying parameter $v$---can only be \emph{indeterminate}; i.e.\ its numerator and denominator must be both zero (the unrealistic alternative of a ratio of infinities can be excluded). Hence $v = c$ \emph{and} $c=1/\sqrt{\Omega}$. Therefore \emph{without any need for postulation}---such as in a recent major book\footnote{A statement in Rindler's recent outstanding work [34, p 15, p 57] `So the only function of the second postulate is to fix the invariant velocity. And Maxwell's theory and the ether-drift experiments clearly suggest that it should be c', appears to be representative of prevailing opinion on the matter. Although the same author also establishes in [17] the existence of a cosmic limit speed from the first postulate alone, this statement overlooks the possibility of an argument such as that above.}---we can draw two important final conclusions from physical observations \emph{in conjunction with the triplet velocity equation}:
\begin{align} 
 & \emph{Every cosmic limit speed instance exhibits its speed in the space–time frame} \nonumber \\  & \emph{of every observer, as well as in the space–time frame of its source.} \end{align}   
\begin{align} 
 & \emph{As far as can be ascertained kinematically from the current limits of} \nonumber \\  & \emph{observational precision, the speed of light is an instance of the cosmic limit speed.} \end{align}

\section{Response time intervals of arbitrary velocity signals}
Should there be a minute---currently undetectable---difference between the limit speed and the speed of electromagnetic wave propagation, this might be ultimately established using the triplet velocity ratio equation to measure, i.e.\ indirectly quantify, $\Omega$, by clocking signal response intervals between inertial vehicles in space. In practice of course our signals would use electromagnetic waves, but in principle any effectively non-intrusive signal carrier which has \emph{arbitrary} but constant velocity $s$ \emph{relative to the space–time frame of its emitter} would suffice---as we shall see, our signal speed need not be even approximately equal to the limit speed (cf Mermin's discussion [23]). Naturally, absence of appreciable gravitational fields would be imperative, which means that no bodies of significant mass should be anywhere near or between the vehicles (at least two) deployed, i.e. the experiments could be successfully performed only in outer space. Accuracy would depend on attainable inter-vehicle speed(s) as well as clock resolution (see also the remark in section 5).

We imagine (figure 7) a space vehicle ($I'$), travelling at a constant unknown velocity $v$ relative to a space base station ($I$) carrying a precision clock, and assume that the line of motion of $I'$ relative to $I$ has an \emph{unknown offset distance} $M$ from $I$.

At base time $T_0$, $I'$ is at \emph{an unknown distance $L$ from the minimum offset point}. For mathematical reasons we adopt a velocity standard such that the speed $s$ of whatever signal is used has the value one. At $T_0$, $I$ transmits at speed $s \equiv 1$ a signal which reaches $I'$ at $t_1$. $I'$ immediately actively transmits back (as opposed to passively mirroring) a signal which reaches $I$ at $T_1$, at speed $s \equiv 1$ \emph{relative to itself in its own $I'$ space–time frame}. These exchanges are continued with each $T_n$ being recorded. Time and distance intervals and velocities are considered primarily from the space–time frame of the base station $I$.

\begin{figure}[t]  
{\centering \includegraphics[width=110mm,height=50mm]{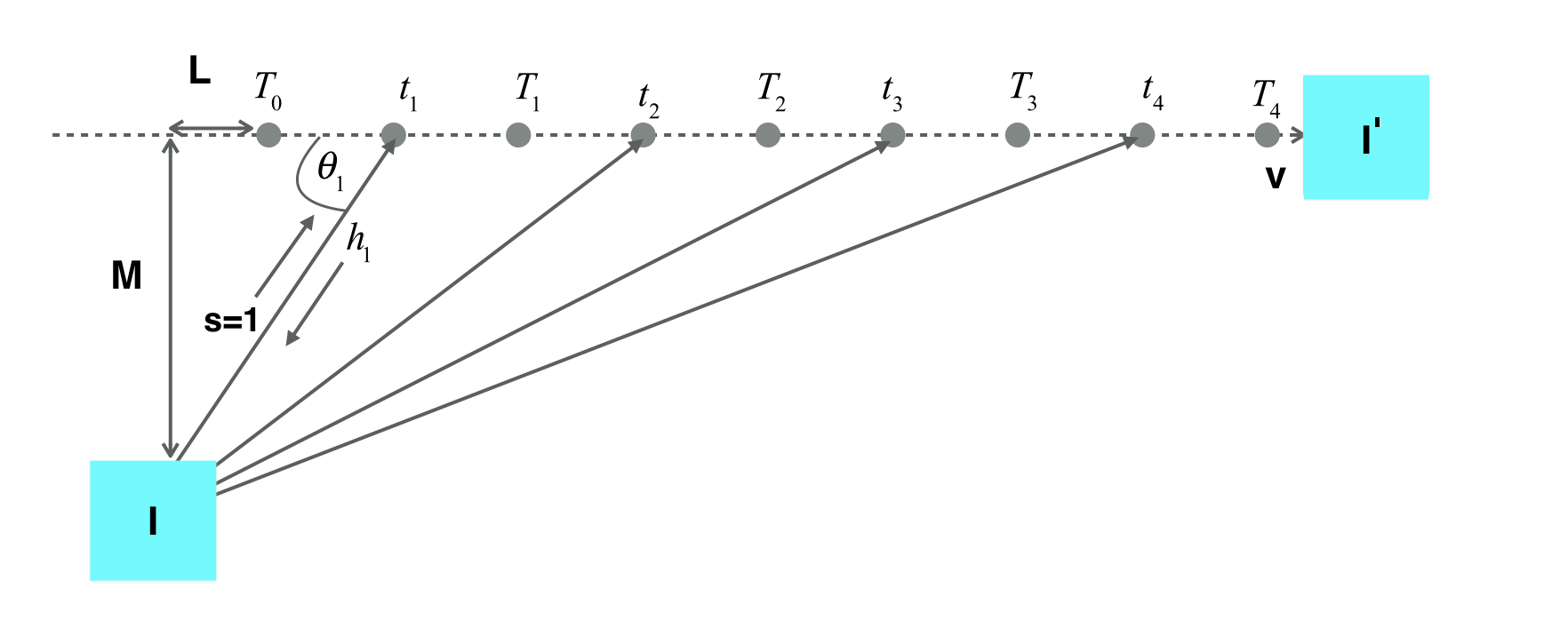} \par} 
\caption{Signal exchanges between non-collinear probes.}\label{fig7}  
 \end{figure}

\subsection{The return signal speed value}
The speed in frame $I'$ of a return signal \emph{along the frame's line of motion} is

\begin{equation} h'_a= \left| \frac{dx'}{dt'} \right|=1. \nonumber\end{equation} 

In frame $I$, from (11) (considering the \emph{wavefront} as a `third frame'),

\begin{equation} h_a= \left| \frac{1-v}{1-v\Omega} \right|. \end{equation} 

The speed of a return signal wave \emph{perpendicular} to the frame's line of motion has \emph{in frame} $I'$ the value 

\begin{equation} h'_p= \left| \frac{dy'}{dt'} \right|=1. \nonumber\end{equation}

For this wavefront's speed,

\begin{equation}  \left| \frac{dx'}{dt'} \right|=0. \nonumber\end{equation}

Because the perpendicular relative velocity is zero, `vertical' length is the same in both frames, i.e.\ $\Delta y =  \Delta y'$. From equation ((22)ii)---inverted (the Lorentz transformation time equation)---we have   

\begin{equation} \Delta t=\frac{\Delta x'v\Omega+\Delta t'}{\sqrt{1-v^2\Omega}} .  \nonumber \end{equation}

In the limit this gives the \emph{time dilation} formula

\begin{equation} \frac{dt}{dt'}=\left(\frac{dx'}{dt'}v\Omega+1\right)/\sqrt{1-v^2\Omega}=\frac{1}{\sqrt{1-v^2\Omega}}  .   \nonumber   \end{equation}

Thus in frame $I$ the emitted \emph{perpendicular} wavefront's speed is

\begin{equation} h_p=\left|\frac{dy}{dt}\right|=\left|\frac{dy'}{dt'}\frac{dt'}{dt}\right|= 1\sqrt{1-v^2\Omega} .\end{equation}

Using the geometry and equations (28) and (29), the speed along each return signal path---
    at angle $\theta_n$ with the line of motion---is given in general for each $n$ by

\begin{equation} 
h_n = \sqrt{(h_a\cos\theta_n)^2+ (h_p\sin\theta_n)^2  }
=\frac{\sqrt{[(1-v)^2/(1-v\Omega)^2](L+vt_n)^2+(1-v^2\Omega)M^2  }}{t_n-T_{n-1}}.
\end{equation} 

\subsection{A recursive signal response time formula}
The outgoing and return signal distances being equal,
\begin{equation} 1(t_n-T_{n-1}) = h_n(T_n-t_n).  \end{equation}

%%%%%%%%%%%%%%\[\color{blue} 1/h_n = \frac{(t_n-T_{n-1})}{\sqrt{[(L+vt_n)(1-v)/(1-v\Omega)]^2+(1-v^2\Omega)M^2   }}  \]
From (30) and (31),
\begin{equation} T_n=t_n+\frac{(t_n-T_{n-1} )^2}{
\sqrt{[(L+vt_n)(1-v)/(1-v\Omega)]^2+(1-v^2\Omega)M^2      }}.
\end{equation}
From the geometry we have directly
\[  (t_n-T_{n-1})^2=M^2+(L+vt_n)^2 \]
%%%%%%%%\[\color{blue}  t_n^{2} - 2t_{n}T_{n-1}  +T_{n-1}^2  =   M^{2}+L^{2}+2Lvt_{n} +v^{2}t_{n}^{2}    \]
%%%%%%%%\[\color{blue}  (1-v^{2})t_n^{2} - 2t_{n}(T_{n-1} +Lv)   =   M^{2}+L^{2} - T_{n-1}^2  \]
%%%%%%%%\[\color{blue} t_n^{2} - \frac{2t_{n}(T_{n-1} +Lv)}{ (1-v^{2})}   +\frac{(T_{n-1}+Lv)^{2}}{ (1-v^{2})^{2}} =   \frac{M^{2}+L^{2}- T_{n-1}^2 }{ (1-v^{2})}  +\frac{T_{n-1}^{2}+2 T_{n-1}Lv+L^{2}v^{2}}{ (1-v^{2})^{2}}\]
%%%%%%%%\[\color{blue}  (t_n-\frac{(T_{n-1}+Lv)}{ (1-v^{2})})^{2}=   \frac{(M^{2}+L^{2})(1+v^{2})- T_{n-1}^2(1-v^{2})+T_{n-1}^2+2LvT_{n-1}+L^{2}v^{2}}{ (1-v^{2})^{2}}   \]
%%%%%%%%\[\color{blue} (t_n-\frac{(T_{n-1}+Lv)}{ (1-v^{2})})^{2} =   \frac{(M^{2}+L^{2})(1-v^{2}) +v^{2}T_{n-1}^2+2LvT_{n-1}+L^{2}v^{2}}{ (1-v^{2})^{2}}   \]
%%%%%%%%
i.e.\ 
\begin{equation} t_n=\frac{T_{n-1}+vL+\sqrt{vT_{n-1}(vT_{n-1}+2L)+G^2 } }{(1-v^2)}   \end{equation}

where $G^2 \equiv L^2 + M^2(1 - v^2)$, and therefore 

\begin{equation} M^2=(G^2-L^2)/(1-v^2).\end{equation}
Hence from (32) and (33) we obtain a general formula for $T_n$ in terms of $T_{n-1}$, $v$, $L$, $G$ and $\Omega$:

\begin{align} & T_n= \frac{T_{n-1}+vL+\sqrt{vT_{n-1}(vT_{n-1}+2L)+G^2     } }{ 1-v^2 }\qquad +
\nonumber \\ &\frac{(1-v\Omega)[v^2T_{n-1}+vL+\sqrt{vT_{n-1}(vT_{n-1}+2L)+G^2 }]^2/(1-v^2)        }
{\sqrt{[vT_{n-1}+L+v\sqrt{vT_{n-1}(vT_{n-1}+2L)+G^2 }]^2(1-v)^2+(1-v^2)(1-v\Omega)^2(1-v^2\Omega)(G^2-L^2)}}
    .  \end{align}

\section{Measuring $\Omega$ using two non-collinear space probes and a single clock}
Assuming $T_0 = 0$ and adopting the $T_1$-interval as \emph{unit time} (i.e. dividing $T_1$ itself, $T_2$, $T3$ and $T_4$ by $T_1$) simplifies equation (35) for n = 1 and 2. Four instances of the equation with known $T_n$ then allow each of the four unknowns $v$, $L$, $G$ and $\Omega$ to be expressed explicitly in turn, as solutions of ratios of polynomials of degree 10, 8, 4 and 3 respectively, whose coefficients in each case contain only the remaining three parameters and measured $T_n$ and $T_{n-1}$. This enables the four unknown parameters to be solved for by convergence methods, the results being in terms of the arbitrary signal speed $s$ as unit velocity and the measured $T_1$-interval as unit time. The rational function solutions were obtained using the Symbolic Maths functions of the MAPLE program (cf Acknowledgments). These equations, which are too long for inclusion in the paper, were presented for referee scrutiny.  \textbf{PDF script now available at}
https://spacetimefundamentals.files.wordpress.com/2020/04/maplescript2003.pdf \\

\noindent \textbf{Remark}. The values of $v$, $L$, $M$ (using equation (34)) and $\Omega$ could be continually monitored by interchange of trains of secondary-indexed signal pulses between the non-collinear probes, with minimal time intervals between consecutive initial pulses. Each convergence computation would use the last four recorded response time intervals between identically secondary-indexed $T_k$. The values of $L$ and $G$ would then be observed to change accordingly each time, but $M$, $v$ and $\Omega$ would remain constant. Of possible interest here is the fact that a transient gravitational wave would \emph{temporarily} distort such computation results, with $M$ and $v$ subsequently having new but again constant values. A triplet of such non-collinearly moving probes would allow three-dimensional spatial and chronological correlation, possibly permitting the rate of propagation of gravitational waves to be quantified, and would involve far less equipment demands than those envisaged for the LISA space project (http://lisa.jpl.nasa.gov) which entails three mutually \emph{stationary} space vehicles. In addition therefore to formally measuring $\Omega$ and confirming to the limits of single-clock precision the numerical identity \emph{or otherwise} between the limit and signal speeds, such an experiment might be a comparatively straightforward method of detecting gravitational waves. Of course the equation solutions are far simpler if $\Omega$ is taken as unity.

\section{Measuring $\Omega$ using three collinear space probes and two clocks}

\footnote{Chapter 6 \emph{Measuring the Cosmic Limit Speed} in [36] establishes equations [40] and [41] \emph{without} using this paper's section 4 recursion formula.}We conclude with a mathematically simpler experiment. Although less feasible due to the difficulty of achieving collinear movement between a triplet of vehicles, it leads to an interesting equation for $\Omega$ which is \emph{explicit} in terms of signal response time interval ratios and the--- arbitrary---speed of \emph{any} signal used. 

\subsection{The consecutive response intervals ratio}
If the frames are in \emph{collinear} motion, i.e.\ $M = 0$ and $G = L$, then the recursive formula (35) reduces to 

\begin{equation} T_n=\frac{(T_{n-1}+L)}{(1-v)}+\frac{(1-v\Omega)(vT_{n-1}+L)}{(1-v^2)}      .\end{equation}
\newpage
Assuming $T_0 = 0$, we obtain the \emph{consecutive response ratio}:

\begin{equation} p\equiv  \frac{T_2-T_1}{T_1}=\frac{(1-v^2\Omega)}{(1-v)^2}\end{equation}

from which we have

\begin{equation} (p+\Omega)v^2-2pv+p-1=0 .\end{equation}

Solving for $v$ (discarding the higher-speed solution),

\begin{equation} v=\frac{p-\sqrt{p-(p-1)\Omega}}{p+\Omega} .\end{equation}

 \begin{figure}[t]  
{\centering \includegraphics[width=90mm,height=50mm]{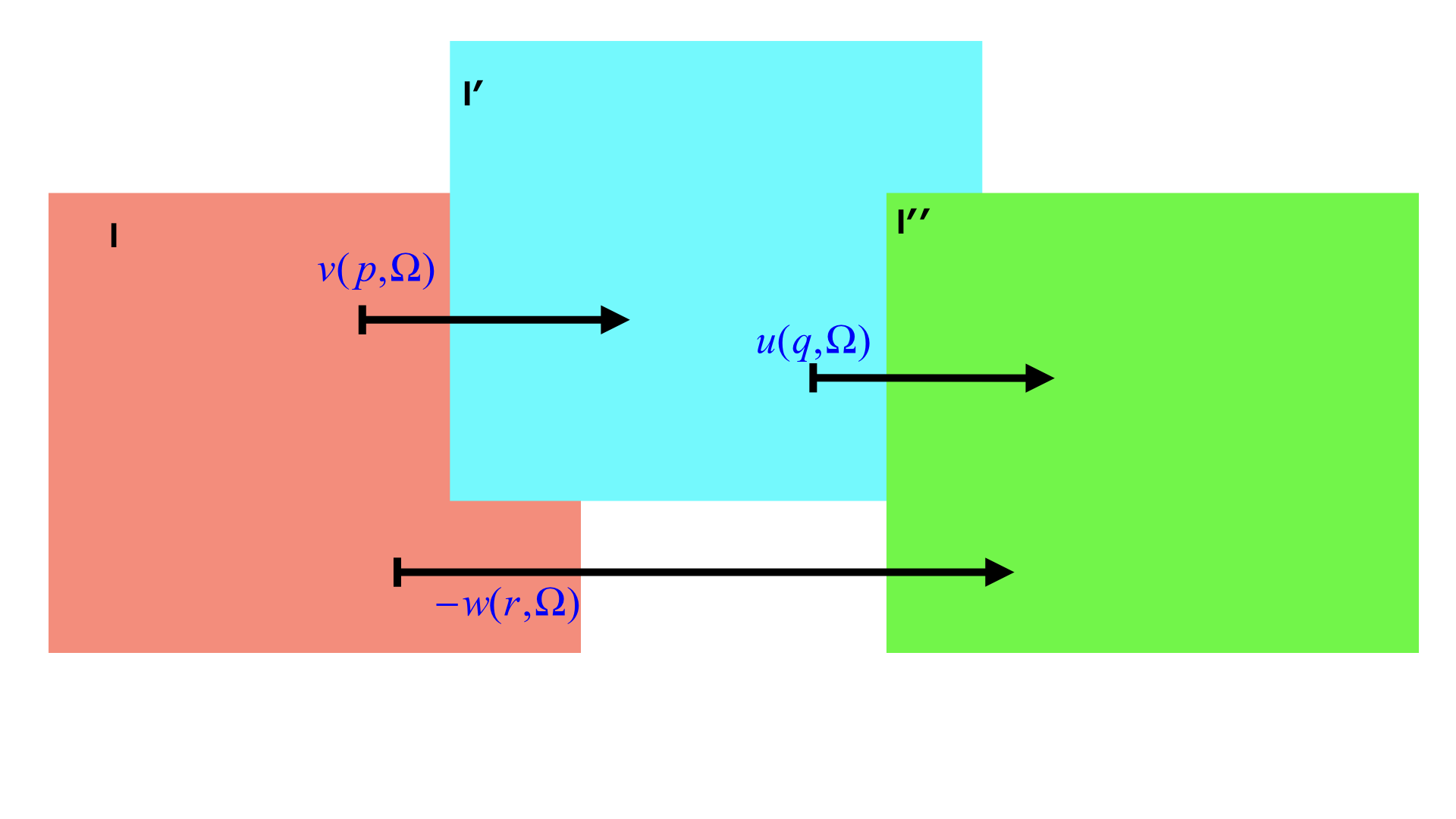} \par} 
\caption{Measuring $\Omega$.}\label{fig8}  
 \end{figure}
%\vspace*{-4mm}
\subsection{A direct quantitative connection between $\Omega$ and the signal unit value $1/s^2$}
A third vehicle (figure 8), frame $I''$ travelling at likewise unknown but constant velocities $u$ and $-w$ relative to frames $I'$ and $I$ respectively and \emph{collinear} with $v$, can return signals to $I'$ which carries an identical clock---not necessarily synchronized---in order to establish the $u$-dependent response intervals ratio $q$. The $-w$-dependent ratio $r$ can be established by signals between $I$ and $I''$ which are initiated and clocked by the base station $I$, removing the need for an $I''$-clock. Using equation (39), we then have for the (unknown) values of $u$ and $-w$,
\begin{equation} u=\frac{q-\sqrt{q-(q-1)\Omega}}{q+\Omega} .\nonumber \end{equation}
\begin{equation} -w=\frac{r-\sqrt{r-(r-1)\Omega}}{r+\Omega} .\nonumber \end{equation}

 The unknown velocities in the triplet velocity equation $\Omega  = (v + u + w)/(-vuw)$ are now replaced by these expressions containing the measured values of $p$, $q$ and $r$, the signal speed $s \equiv 1$ and the unknown $\Omega$:
 \begin{align}  &  \Omega= \frac{ {\frac{{p - \sqrt {p - \left( {p - 1} \right){\Omega}} }}{{p +\Omega}} + \frac{{q - \sqrt {q - \left( {q - 1} \right){\Omega}} }}{{q + \Omega}} - \frac{{r - \sqrt {r - \left( {r - 1} \right){\Omega}} }}{{r + \Omega}}}     }{ {\frac{{p - \sqrt {p - \left( {p - 1} \right)\Omega} }}{{p + \Omega}} \cdot \frac{{q - \sqrt {q - \left( {q - 1} \right)\Omega} }}{{q + \Omega}} \cdot \frac{{r - \sqrt {r - \left( {r - 1} \right)\Omega} }}{{r + \Omega}}}  }.\end{align}
The \emph{explicit} solution of equation (40) for   has been obtained likewise with the help of the MAPLE program\footnote{Solution (41) was produced with the help of the Symbolic Maths MAPLE program, but only after transformation of equation (40) (by Kevin Hutchinson---see Acknowledgments) to a set of four simpler equations with radicals replaced by single variables. The emergent `RootOf' expressions were then further processed leading to the solution which is one of two, the other being identical except for the sign of the $\sqrt{pqr}$ term (see equation (39)) and producing the same value but involving in practice unattainably large values of $p$, $q$ and $r$ (i.e. vehicle velocities $v$, $u$ and $w$ close to $c$).}: 
\begin{align} &\Omega=  \frac{{[pq(r + 1) - 2(p + q)r]pq(r - 1) + {{\left[ {(p - q)r} \right]}^2}(r + 1)/(r - 1)}}{{\left\{ {{{\left[ {(p - q)r} \right]}^2} + pq(r - 1)[pq(r - 1) - 2(p + q - 2)r]} \right\}}}  \,\,\, +\nonumber \\  & \,\, \,\, \frac{ {4\sqrt {pqr} [pq(r + 1) - (p + q)r][{{\left[ {pq(r - 1)} \right]}^2} - {{\left[ {(p - q)r} \right]}^2}] - \frac{{2{{[{{\left[ {pq(r - 1)} \right]}^2} - {{\left[ {(p - q)r} \right]}^2}]}^2}}}{(r - 1)}}  }  
 {{{{\left\{ {{{\left[ {(p - q)r} \right]}^2} + pq(r - 1)[pq(r - 1) - 2(p + q - 2)r} \right\}}^2}}} .\end{align}
If the signal speed $s$ happens to be equal to the limit speed $1/\sqrt{\Omega}$ , i.e.\ $\Omega=1/s^2   \equiv 1$, then (40) and (41) would reduce to $r = pq$ and $\Omega  = 1$ respectively. Values where $r = pq$ would therefore establish the signal velocity $s$ as the cosmic limit speed.

\section{Summary}
The universal constant $\Omega$, from dynamics the mass-to-energy ratio, emerges `naturally' from elementary kinematics as the universal proportionality factor between space–time chronocity and velocity as well as between the cyclic sum and negative product of triplet interframe velocities. Its inverse square root, by virtue of velocity addition uniqueness the cosmic limit speed, is directly quantified by one of the latter's logically established instances---the rate of electromagnetic wave propagation. $\Omega$ is also measurable independently---at least in principle---of such a limit speed signal, and moreover can be expressed explicitly as a function of signal response time ratios and the emitter-relative constant speed of any suitable non- intrusive signal. This contrasts with the still current prevalence of the second postulate as a cornerstone of special---and general---relativity.

\section*{Acknowledgments}
The author is indebted to Ian Elliott of Dunsink Observatory (Dublin Institute for Advanced Studies) for ongoing guidance, to David Mermin of Cornell University for constructive criticisms of an (unpublished) August 2001 version of the paper which corresponded to present sections 1-3, to Liam Little for initial inspiration and steadfast support, to Kevin Hutchinson of the University College Dublin Mathematics Department for assistance in obtaining an explicit solution of the key variable, and to Annraoi de Paor of the UCD Electronic and Electrical Engineering Department for useful comments. Introduced through Nikolai Demidenko and Sergei Kucherenko of Guildford, UK, Sergei Yakovlenko, head of the Moscow General Physics Institute, Kinetics Department, very kindly provided information on Ignatowski. Special tribute is also due to the developers of the MAPLE program (Waterloo Maple Inc., Ontario, Canada) for the power, convenience and inspiration it afforded even to a non-mathematician, and to Adept Scientific, UK, for generously providing it at an academic price for the purposes of this paper.

The following libraries/institutions allowed generous access to books and papers: Dublin Institute for Advanced Studies (Theoretical Physics), Trinity College, Dublin (FitzGerald, Hamilton), University College, Dublin (Belfield), Queen's University, Belfast, University of Cambridge (Applied Mathematics and Theoretical Physics, St John's), Oxford University (Radcliffe), Universit\"{a}t Erlangen-N\"{u}rnberg, Max-Planck-Institut, Munich, Humboldt Universit\"{a}t, Berlin, Leopold-Franzens-Universit\"{a}t, Innsbruck, Massachusetts Institute of Technology (Hayden Memorial), Boston.

The author has a degree in electrical engineering (University College Dublin, 1968), and a background in artificial intelligence systems for process control (Germany).

\end{document}